\documentclass[prb,aps,twocolumn,draft,tightenlines,%
showpacs,floatfix]{revtex4}
\usepackage{psfig}
\usepackage{amssymb}
\usepackage{psfig}
\usepackage{amsmath}
\usepackage{colordvi}
\begin{document}

\title{Spin-Hall effect in two-dimensional mesoscopic hole systems}
\author{M. W. Wu}
%\thanks{Author to whom all correspondence should be addressed}%
\email{mwwu@ustc.edu.cn}
\author{J. Zhou}
\affiliation{Hefei National Laboratory for Physical Sciences at
  Microscale, University of Science and Technology of China,
Hefei, Anhui, 230026, China and\\
Department of Physics, University of Science and
Technology of China, Hefei, Anhui, 230026, China}
\altaffiliation{Mailing Address.}
\date{\today}

\begin{abstract}
The spin Hall effect in two dimensional hole systems is studied
by using the four-terminal
Landauer-B\"{u}ttiker formula with the help of Green functions.
We show that the heavy (light) hole spin Hall effect exists even when
there are {\em no} correlations
between the spin-up and -down heavy (light) holes  and when
the $\Gamma$-point
degeneracy of the heavy hole and light hole bands is lifted
due to the confinement or recovered by the strain.
When only a heavy hole charge current without any spin polarization
is injected from one lead, under right choice of lead voltages,
one can get a pure heavy- (light-) hole spin current,
combined with a possible impure light (heavy) hole spin current
from two transverse leads.
The spin Hall coefficients of both heavy and light holes
depend on the Fermi energy, device size and the disorder
strength. It is also shown that
the spin Hall effect of two dimensional hole systems
is much more robust than that of electron systems with
the  Rashba spin-orbit coupling
and the spin Hall coefficients do not decrease with the system
size but tend to some
nonzero values when the disorder strength is smaller than some critical value.
\end{abstract}

\maketitle

Recently  spintronics has become one of the major focuses of the
condensed matter physics.\cite{prinz} Generating and measuring
spin current is still a primary task of this field. The spin Hall
effect (SHE) is considered to be a hopeful way to produce the spin
current. This effect, originally proposed as the extrinsic SHE as
it requires spin-dependent impurity scattering, was  first studied
by Dyakonov and Perel'\cite{dyak} in the early seventies and
lately by Hirsch\cite{hirsch} and  Zhang.\cite{shufeng} It
further evolves into the  intrinsic SHE very recently as it
appears  in  clean systems  with the intrinsic spin-orbit (SO)
coupling, such as two-dimensional electron systems
(2DES)\cite{sinova,hu,rashba} and bulk hole systems\cite{SCzhang}
without any impurity scattering.

In 2DES, the intrinsic SO coupling comes from the
Rashba\cite{ras} or Dresselhaus\cite{dress} terms which
correlate the two spin states of electrons and lead to
nonzero spin correlations $\langle a^\dagger_{{\bf k}\sigma} a_{{\bf k}
-\sigma}\rangle$.
Calculation based on  Kubo formula
found that  the spin Hall coefficient (SHC) is a universal
value $e/8\pi$ in macroscopic  2DES with the Rashba SO coupling.\cite{sinova}
The fate of the intrinsic  SHE
in the presence of disorder is an important problem
and has raised a lot of controversies.
On one hand, Burkov {\em et al.}\cite{burkov} and
Schliemann and  Loss\cite{loss1} pointed
out that the SHE only survives at weak disorder scattering in
macroscopic 2DES. On the other hand,
more analytical and numerical calculations\cite{mish, inoue,sheng1,lei}
showed that the SHE
 vanishes even for any weak disorder.

Situations are quite different in the mesoscopic systems: using
the four-terminal Landauer-B\"uttiker formula,
Sheng {\em et al.}\cite{sheng} and Nikoli\'{c} {\em et al.}\cite{nikolic} showed that
the SHC does not take a universal value in a finite mesoscopic 2DES,
but depends on the magnitude of the SO coupling, the
electron Fermi energy, and the disorder strength. Moreover,
Sheng {\em et al.} showed that when the disorder is smaller than some critical
values, the SHC does not decrease
with system size but instead goes to some nonzero values.
This robustness to the disorder in mesoscopic systems is suggested from the
effect of the boundary effects (leads).\cite{sheng1}
Therefore a pure spin current (no charge current associated with the
spin current) can be
induced by unpolarized charge current due to SHE in mesoscopic
2DES.\cite{pareek,sheng,nikolic}

The SHE in macroscopic hole systems has also been studied.
Murakami {\em et al.}\cite{SCzhang} predicted that the
pure spin current can be obtained in $p$-type bulk  semiconductors.
They further suggested that this SHE comes
 from the Dirac magnetic mono-pole in momentum space
because of the fourfold degeneracy  at the
$\Gamma$-point of the valence band. Then
Murakami\cite{murakami} pointed out that the vortex correction
that was reported to kill the SHE in 2DES\cite{inoue}
is identically zero in the Luttinger model so that the SHE can survive
the impurity scattering. This is verified by numerical calculation
in bulk lattice Luttinger model very recently.\cite{chen} On the other hand,
Schliemann and Loss\cite{loss} and Bernevig and
Zhang\cite{zhang} studied the SHE in 2D
hole gases in which the HH  and the LH bands are
no longer degenerate at the $\Gamma$-point (no mono-pole in this case).
However, they considered the
case where there are still correlations between spin-up and -down HH's
by introducing {\em additionally} the Rashba SO coupling in the hole systems.
Therefore, whether the magnetic mono-pole is crucial for the
SHE in hole systems is still unknown.

We notice that the study of the SHE in mesoscopic 2D hole systems is
still absent. There are a lot of differences
between 2DES and 2D hole systems: In 2DES, spin-up and -down
electrons are always correlated via the Rashba or the Dresselhaus SO couplings.
Nevertheless, in 2D hole systems the situation becomes much more
complicated: For (001)  quantum wells (QW's) with small well width where
only the lowest subband is important,  unless one adds an additional
Rashba spin-orbit coupling, there are {\em no}
direct or indirect spin correlations
between spin-up and -down HH's (LH's), {\em i.e.},
$\langle a^\dagger_{{\bf k}\sigma}a_{{\bf k}-\sigma}\rangle\equiv0$
with $\sigma=\pm\frac{3}{2}$ for HH's and $\pm \frac{1}{2}$ for LH's.
The spin-up HH's (LH's) are only
coupled with the spin-down LH's (HH's). Adding strain can change the
relative positions of the HH and  LH bands and the $\Gamma$-point degeneracy
can be recovered. It is interesting to
see  whether there is still SHE in the absence of
any correlations between spin-up and -down HH's (LH's) and it
is also interesting to see  the
role of magnetic mono-pole to the SHE in the hole systems.
In this paper, we study these problems
by using the four-terminal
Landauer-B\"{u}ttiker formula with the help of Green functions.% We find
%even though there are {\em no}  correlations between the spin-up  and
%-down HH's (LH's),  surprisingly the SHE still exists.
%Significantly, we obtain {\em pure} spin currents of
%HH  and  {\em impure} spin currents of LH at the same time.
%When the condition changes, we can also get  pure spin currents of LH and
%impure spin currents of HH.

We consider a 2D hole system which is consisted of a square
conductor of width $L$ attached with four ideal leads  of width
$L/2$ without any spin-orbit coupling (as illustrated in Fig.\ 1)
in a (001) QW of width $a$. Due to the confinement of the QW, the
momentum states along the growth direction ($z$-axis)
 are quantized. We take only the lowest subband for small well width.
Then the  Luttinger Hamiltonian in the momentum space reads with
the matrix elements arranged in the order of $J_{z}=\frac{3}{2}$,
$\frac{1}{2}$, $-\frac{1}{2}$ and $-\frac{3}{2}$:\cite{trebin}
\begin{equation} H =\frac {1}
{2m_0}\left(\begin{array}{cccc}
P+Q & 0 & R & 0 \\
0 & P-Q & 0 & R \\
R^\dagger & 0 & P-Q & 0 \\
0 & R^\dagger & 0 & P+Q
\end{array}\right)
\end{equation}
where $P\pm Q=(\gamma_1\pm\gamma_2)(P_x^2+P_y^2)+E_0^{\pm}$, and
$R=-\sqrt{3}[\gamma_2(P_x^2-P_y^2)-2i\gamma_{3}P_{x}P_{y}]$.
It is noted that the off-diagonal terms $R$ and $R^\dagger$
mix the HH with the LH. In
real space the Luttinger Hamiltonian can be written in the tight-binding version as:
\begin{eqnarray}
 &&  H=\sum_{i,j,\sigma=\pm\frac{1}{2}\pm\frac{3}{2}}
[E_0^{\pm}-(\gamma_1\pm\gamma_2)4t+\epsilon_{i,j}]a_{i,j,\sigma}
^{\dag}a_{i,j,\sigma} \nonumber\\
& &+\sum_{\begin{subarray}{c} i,j,\delta=\pm1 \\ \sigma=\pm\frac{1}{2},\pm\frac{3}{2}
\end{subarray}}
(\gamma_1\pm\gamma_2)t[a_{i+\delta,j,\sigma}^{\dagger} a_{i,j,\sigma}+a_{i,j+\delta,
    \sigma}^{\dagger}a_{i,j,\sigma}] \nonumber \\
 &&+\{\sum_{\begin{subarray}{c} i,j,\delta=\pm1 \\ \lambda=0,1\end{subarray}}
[(-\sqrt{3})\gamma_{2}t(a_{i+\delta,j,\frac{3}{2}-\lambda}^{\dagger}
a_{i,j,-\frac{1}{2}-\lambda}\nonumber\\
  &&-a_{i,j+\delta,\frac{3}{2}-\lambda}^{\dagger}a_{i,j,-\frac{1}{2}-\lambda})
% \nonumber \\
+\frac{\sqrt{3}}{2}i\gamma_{3}t(a_{i+\delta,j+\delta,\frac{3}{2}-\lambda}^{\dagger}
  a_{i,j,-\frac{1}{2}-\lambda}\nonumber \\
  &&-a_{i+\delta,j-\delta,\frac{3}{2}-\lambda}^{\dagger}a_{i,j,-\frac{1}{2}-\lambda})]+H.C. \}\ ,
\end{eqnarray}
where $i$ and $j$ denote the coordinates along the $x$- and $y$-axis;
$\gamma_1$, $\gamma_2$ and $\gamma_3$ are the Luttinger coefficients,
and $m_0/(\gamma_1\pm\gamma_2)$ are the
effective masses of the HH and the LH in the $x$-$y$ plane
with $m_0$ representing  the free electron mass;
 $t=-\hbar^{2}/2m_{0}{a_{0}^{2}}$ is the unit of
energy with $a_{0}$ standing for the ``lattice'' constant,
and $(\gamma_1\pm\gamma_2)t$ stands for the hopping energy;
$E_0^{\pm}=(\gamma_1\mp 2\gamma_2)\frac{\pi^2}{a^2}|t|$ is the first
subband energy in the $z$ direction;
$\epsilon_{i,j}$ accounts for the spin-independent disorder,
 which is a random value in the range $[-W/2,W/2]$.
Additionally
\begin{equation}
H_{strain}=\sum_{i,j,\sigma=\pm\frac{3}{2},\pm\frac{1}{2}}
\epsilon_{|\sigma|}^{s}a_{i,j,\sigma}^{\dagger}a_{i,j,\sigma}
\end{equation}
is the strain  Hamiltonian where $\epsilon_{|\sigma|}^{s}$ is the strain-induced
energy with $\epsilon_{\frac{3}{2}}^s\not=\epsilon_{\frac{1}{2}}^s$.\cite{strain}
Therefore, by adding strain, one may either further increase the separation
between the HH and the LH or reduce it to recover the $\Gamma$-point degeneracy.
Moreover, it is seen from the Hamiltonian that there is not any direct or indirect spin flip
between the spin-up and -down HH 's or between the spin-up and -down
LH's in the Hamiltonian.

\begin{figure}[htb]
\vskip-0.3cm
\centerline{\psfig{figure=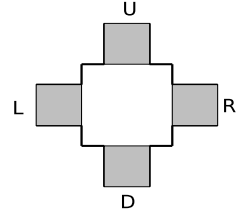,width=3cm,height=3cm,angle=0}}
\vskip-0.3cm
  \caption{Schematic view of a four-terminal junction
of width $L$ attached with
four semi-infinite leads without any spin orbit coupling of width $L/2$.}
\end{figure}

The spin dependent transmission coefficient
from $\mu$ terminal with spin $\sigma$
to $\nu$ terminal with spin $\sigma^\prime$ is calculated using
the Green function method\cite{Da}
$
T^{\sigma \sigma^\prime}_{\mu\nu}=\mbox{Tr}[\Gamma^{\sigma}_{\mu}
G^{\sigma\sigma^\prime R}_{\mu\nu}
\Gamma^{\sigma^\prime}_{\nu}G^{\sigma
^\prime\sigma A}_{\nu\mu}]$,
in which $\Gamma_{\mu}=i[\Sigma^{R}_{\mu}-\Sigma^{A}_{\mu}]$
represents the self-energy function for the isolated ideal leads.\cite{Da}
We choose the  perfect ideal Ohmic contact between the
leads and the semiconductor. $G^{\sigma\sigma^\prime R}_{\mu\nu}$ and
$G^{\sigma\sigma^\prime A}_{\nu\mu}$,  which can be obtained from
$G^{R(A)}=(E-H_{C}-\Sigma^{R(A)})^{-1}$,
are the retarded and advanced Green functions
for the conductor, but with the effect from the leads included.
Here $\Sigma^{R(A)}$ represents the sum of the
retarded (advanced) self-energies of the four leads.

\begin{figure}[htb]
\vskip-0.3cm
 %\centering
  \psfig{figure=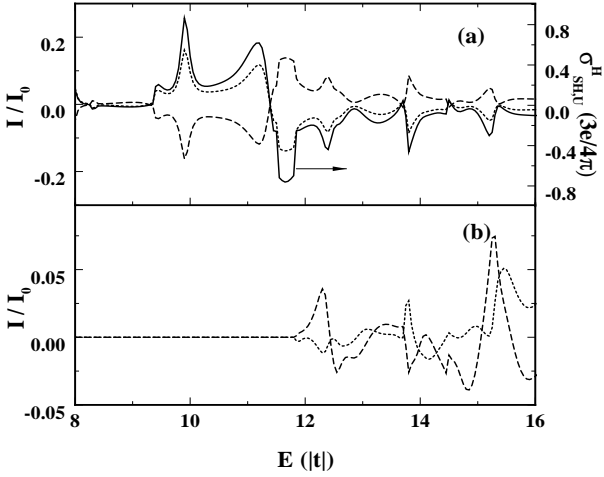,width=8cm,height=7.6cm,angle=0}
\vskip-1cm
  \caption{Particle currents
and SHC {\em v.s.} the Fermi energy $E$ when the pure HH spin current is
obtained. $L=12$.
(a) Solid curve: $\sigma_{SH,U}^{H}$;
Dotted curve: $I_{U}^{\frac{3}{2}}(=I_{D}^{-\frac{3}{2}})$;
Dashed curve: $I_{U}^{-\frac{3}{2}}(=I_{D}^{\frac{3}{2}})$.
(b) Dotted curve: $I_{U}^{\frac{1}{2}}(=I_{D}^{-\frac{1}{2}})$;
Dashed curve: $I_{U}^{-\frac{1}{2}}(=I_{D}^{\frac{1}{2}})$.
}
\end{figure}

\begin{figure}[htb]
\vskip-0.3cm
\psfig{figure=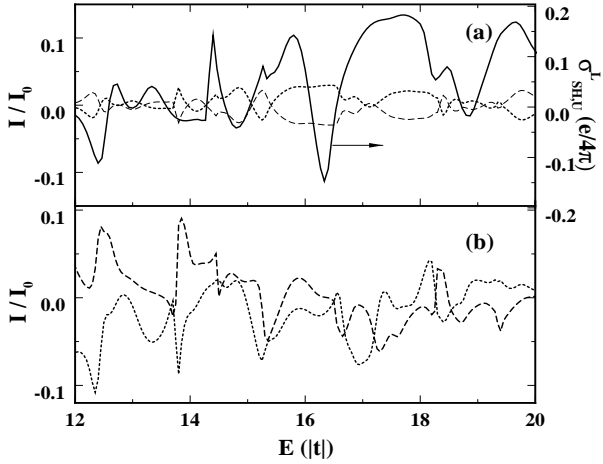,width=8cm,height=7.6cm,angle=0}
\vskip-1cm
\caption{Particle currents
and SHC {\em v.s.} the Fermi energy $E$ when the pure LH spin current is
obtained. $L=12$.
(a) Solid curve: $\sigma_{SH,U}^{L}$;
Dotted curve: $I_{U}^{\frac{1}{2}}(=I_{D}^{-\frac{1}{2}})$;
Dashed curve: $I_{U}^{-\frac{1}{2}}(=I_{D}^{\frac{1}{2}})$.
(b) Dotted curve: $I_{U}^{\frac{3}{2}}(=I_{D}^{-\frac{3}{2}})$;
Dashed curve: $I_{U}^{-\frac{3}{2}}(=I_{D}^{\frac{3}{2}})$.
}
\end{figure}

We perform a numerical calculation when the leads are connected to
isolated reservoirs at chemical potentials $E+V_\mu$ ($\mu=L$, $U$, $R$
and $D$), with the differences between each other  caused
by the applied gate voltages.
The particle current going through the lead $\mu$ with
spin $\sigma$ can be obtained by the
Landauer-B\"{u}ttiker formula\cite{Bu}
$I_{\mu}^{\sigma}=\frac{e}{h}\sum_{\nu\ne\mu,\sigma^\prime}[T_{\nu\mu}^{\sigma^\prime\sigma}V_{\mu}-
T_{\mu\nu}^{\sigma\sigma^\prime}V_{\nu}]$.
 The spin current is defined as
 $I_{s,\mu}^{H}=\frac{3\hbar}{2}(I_{\mu}^{\frac{3}{2}}-
I_{\mu}^{-\frac{3}{2}})$ for  HH's and $I_{s,\mu}^{L}
=\frac{\hbar}{2}(I_{\mu}^{\frac{1}{2}}
-I_{\mu}^{-\frac{1}{2}})$ for  LH's in  the
 lead $\mu$ and the SHC  for the pure spin current is defined as
\begin{equation}
\sigma_{SH,\mu}^{H(L)}=I_{s,\mu}^{H(L)}/(V_{L}-V_{R})\ .
\end{equation}
When the charge current of the
hole $I_{h,\mu}^{H(L)}=e[I_{\mu}^{\frac{3}{2}(\frac{1}{2})}+
I_{\mu}^{-\frac{3}{2}(-\frac{1}{2})}]=0$ and the spin current
$I_{s,\mu}^{H(L)}\not=0$ for the lead $\mu$, then
 $I_{s,\mu}^{H(L)}$ is a pure spin current,
otherwise an impure one.

We drive  a unit HH charge current without any spin polarization
into the left lead $L$ ($I_{L}^{H}=I_{0}$ and $I_{L}^{L}=0$).
In order to get
pure spin currents of HH in the upper and down leads
($U$ and $D$), one needs to find a set of suitable combinations of $V_{\mu}$,
which lead to $I_{h,U}^{H}=I_{h,D}^{H}=0$.
This can be obtained by choosing $V_R=0$ for convenience, $V_U=V_D$ due to the
symmetry and $V_{U}/V_{L}=(T^{\frac{3}{2}\frac{3}{2}}_{UL}+T^{-\frac{3}{2}
-\frac{3}{2}}_{UL})/(T_{1}+T_{2})$ with
$T_{1}=T^{\frac{3}{2}\frac{3}{2}}_{LU}+T^{\frac{3}{2}\frac{3}{2}}_{RU}
+T^{-\frac{1}{2}\frac{3}{2}}_{RU}$ and
$T_{2}=T^{-\frac{3}{2}-\frac{3}{2}}_{LU}+T^{-\frac{3}{2}
-\frac{3}{2}}_{RU}+T^{\frac{1}{2}-\frac{3}{2}}_{RU}$.
And one obtains a pure HH spin current when
$I_{s,U}^{H}=\frac{3e}{4\pi}2V_L
(T_1T^{-\frac{3}{2}-\frac{3}{2}}_{UL}-T_2T^{\frac{3}{2}\frac{3}{2}}_{UL})/
{(T_1+T_2)}\not=0$.
This relation can be satisfied thanks to the
phase shift provided by the last term of Hamiltonian (1)
%[{\em i.e.}, $iP_{x}P_{y}$ in Hamiltonian (1)]
when holes hop from site $(i,j)$
to site  $(i+1,j\pm1)$. Similarly one can obtain the pure LH spin current.
The main results of our calculation are
summarized in Figs.\ 2-4.

In Fig.\ 2 a pure HH spin current is generated in the $U$
and $D$ leads of a $12\times12$  square conductor made from
the unstrained QW, where the $\Gamma$-point degeneracy of
the HH and LH is lifted.
 In Fig.\ 2(a), all the particle currents in these two leads
are plotted as  functions of the Fermi
energy. It is seen that $I_{U}^{\frac{3}{2}}=I_{D}^{-\frac{3}{2}}
=-I_{U}^{-\frac{3}{2}}=-I_{D}^{\frac{3}{2}}$, leading
to $I_{s,U}^{H}=-I_{s,D}^{H}$.
Therefore pure HH spin currents with opposite spin polarizations are obtained
in the absence of any  correlations between the spin-up and -down HH's and
the $\Gamma$-point degeneracy.
The SHC $\sigma_{SH,U}^H$ of the upper lead is also plotted in the
same panel. One finds that the  SHC  depends strongly on the Fermi
energy. It is interesting to see that when the energy $E$ is high
enough, LH's   produce an impure spin current as
shown in Fig.\ 2(b), where
$I_{U}^{\frac{1}{2}}=I_{D}^{-\frac{1}{2}}
\ne -I_{U}^{-\frac{1}{2}}=-I_{D}^{\frac{1}{2}}$ and
hence $I_{h,U}^{L}=I_{h,D}^{L}\ne 0$.
One can also obtain a pure LH spin current combined with
an impure HH spin current in both $U$ and $D$ leads when
a unit HH charge current
without any spin polarization injected into the left lead $L$,
by using the condition $I_{h,U}^{L}=I_{h,D}^{L}=0$ as shown
in Fig.\ 3 for the same conductor.  It is seen from Fig.\ 3(a) that
$I_{U}^{\frac{1}{2}}=I_{D}^{-\frac{1}{2}}
=-I_{U}^{-\frac{1}{2}}=-I_{D}^{\frac{1}{2}}$,
which result in the pure LH spin currents with
opposite spin  polarizations  in the $U$ and $D$ leads.
The SHC of the LH again varies with the Fermi energy.
Moreover, from Fig.\ 3(b) one has $I_{U}^{\frac{3}{2}}=I_{D}^{-\frac{3}{2}}
\ne -I_{U}^{-\frac{3}{2}}=-I_{D}^{\frac{3}{2}}$ which lead to an
impure HH spin current with
$I_{h,U}^{H}=I_{h,D}^{H}\ne 0$.

Therefore, both HH and LH pure spin currents
can be obtained through the suitable combination  of the applied voltages.
This provides us a unique way to manipulate the hole spin currents.
Moreover, if one further allows only  the HH charge current
through the right lead $R$, then  pure spin currents
of HH and LH can be obtained at the same time in the $U$ and $D$ leads.

\begin{figure}[htb]
\vskip-0.3cm
\psfig{figure=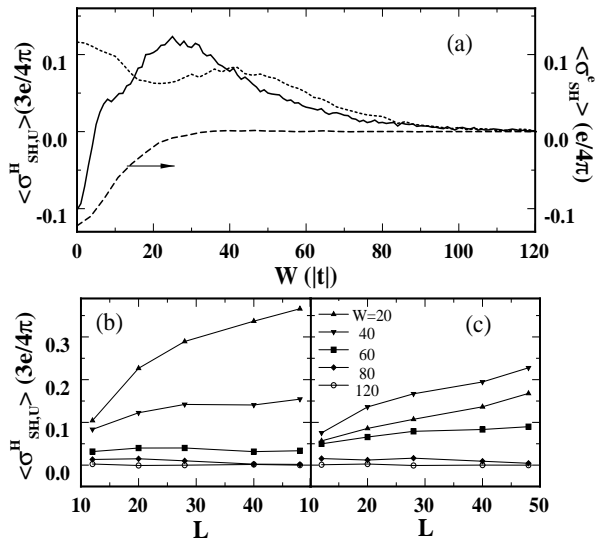,width=8cm,height=8cm,angle=0}
\vskip-1cm
\caption{(a) SHC average over 5000 random disorder configurations versus the disorder strength.
Solid curve: SHC of HH without strain when $E=16|t|$;
Dotted curve: with strain when $E=20|t|$; Dashed curve: SHC of
electron system. $L=12$.
(b) SHC averaged over 1000 random disorder configurations versus $L$ without strain
for $W=20|t|$,
$40|t|$, $80|t|$, and $120|t|$.
(c) same with (b) but with strain.}
\end{figure}

In order to check the robustness of the SHC of the hole system, we
plot in Fig.\ 4(a) the HH SHC of the $U$ lead of a $12\times12$ conductor
with and without strain versus the  strength of disorder $W$
 over $5000$ random disorder configurations. The strain is chosen  to
recover the $\Gamma$-point degeneracy of the HH and LH bands (and thus
the magnetic mono-pole).
For the strain free and strain applied cases, $E=16|t|$ and
$20|t|$ to ensure that the energy always sits around 1/4 of the HH
bandwidth from the $\Gamma$-point.
It is seen that for strain free case,
the SHC is negative when $W$ is small and then becomes positive. It reaches
to the maximum near $W=30|t|$, and then decreases to zero when
$W$ is larger than  $100|t|$ .
When the $\Gamma$-point degeneracy is recovered by the strain,
the SHC is killed by the same disorder strength,
which shows that the degeneracy does not change the robustness of the
SHE. It is noted that there is no correlations between the spin-up and
-down HH's in both cases. We further compare
the robustness of the SHC of 2D hole systems with 2DES.
In Fig.\ 4(a) we also plot the SHC from the Rashba
SO coupling of electrons in a conductor of in the same size as holes
versus the disorder strength
over $5000$ random disorder configurations. As the effective masses of
electron and HH are different, we rescale the energy of electrons by
$\gamma_1+\gamma_2$ in the figure. The energy is again chosen
to be at the 1/4 of the band width from the bottom. The Rashba SO coupling
 $t^{R}_{SO}=0.1|t|$ (before the rescale) is taken to be the
maximum value which can be obtained in the experiment.\cite{nitta}
It is seen that the SHC of electron system approaches to zero at $W=30|t|$
(after rescale), much smaller than the value of HH's.
Therefore the SHC of hole systems is much more robust than electron ones.
This origins from the stronger intrinsic SO coupling of the hole systems.
We further examine the size effect by
calculating the SHC for the  HH
as a function of the system size $L$
under different disorder strengths  in the cases with and without strain.
As seen in  Figs.\ 4(b) and (c), similar to the case
of electron systems with the Rashba SO coupling,\cite{sheng}  the SHC does not decrease
with the  size but goes to some
nonzero value when the disorder strength is smaller than
some critical value. However, when the disorder is large enough, then
the SHC goes to zero with the size. We have also calculated the SHC of LH's
and come to the same conclusion.

In summary, we have performed a mesoscopic investigation of the
SHE for holes in four-terminal (001) QW's with a small well
width so that only the lowest subband is populated, the
 correlations between the spin-up and -down HH's (or LH's) are totally
absent and the $\Gamma$-point degeneracy between the HH and LH bands
are lifted unless a certain strain is applied.
We find that the SHE still exists. Moreover  a pure HH  (LH) spin current
can be generated combined with a possible impure LH  (HH) spin current,
when a HH charge current without any spin polarization is injected
from one lead. The SHC's for both HH and LH
depend on the Fermi energy, device size and the disorder strength.
We also find that the SHE of holes does not need the mono-pole
from the $\Gamma$-point degeneracy.
We show that the SHE of 2D hole systems are much more robust
than that in 2DES. This is consistent with the latest
study in  bulk systems.\cite{chen}
The SHC does not decrease with the system size
 but tends to some nonzero value
 when the disorder strength is smaller than some critical value,
similar to the electron case but where there are direct correlations
between the spin-up and -down electrons.

This work was supported by the Natural Science Foundation of China
under Grant Nos. 90303012 and 10247002, the Natural Science Foundation
of Anhui Province under Grant No. 050460203 and SRFDP.


\begin{thebibliography}{10}

\bibitem{prinz} %G.A. Prinz, Phys. Today {\bf 48}, 58 (1995); Science
%  {\bf 282}, 1660 (1998);
%D. Loss and D. P. DiVincenzo, Phys. Rev. A {\bf 57}, 120 (1998);
{\em Semiconductor Spintronics and Quantum
  Computation}, eds. D. D. Awschalom, D. Loss, and N. Samarth
  (Springer, Berlin, 2002); I. \v Zuti\'c, J. Fabian, and S. Das Sarma,
Rev. Mod. Phys. {\bf 76}, 323 (2004).
\bibitem{dyak}M. I. D'Yakonov and V. I. Perel', Phys. Lett.A {\bf 35}, 459 (1971).
\bibitem{hirsch}J. Hirsch, Phys. Rev. Lett. {\bf 83}, 1843 (1999).
\bibitem{shufeng}Shufeng Zhang, Phys. Rev. Lett. {\bf 85}, 393 (2000).
\bibitem{sinova}J. Sinova, D. Culcer, Q. Niu, N. A. Sinitsyn, T. Jungwirth, and A. H. MacDonald,
  Phys. Rev. Lett. {\bf 92}, 126603 (2004).
\bibitem{hu}J. Hu, B. A. Bernevig, and C. Wu, cond-mat/0310093; S. Q. Shen, Phys. Rev. B {\bf 70}, 081311 (2004).
\bibitem{rashba}E. I. Rashba, Phys. Rev. B {\bf 68}, 241315 (2003); {\em ibid.}
{\bf 70}, 161201 (2004);  {\em ibid.}
{\bf 70}, 201309 (2004).
\bibitem{SCzhang}S. Murakami, N. Nagaosa, and S.C. Zhang,
Science {\bf 301}, 1348 (2003).
\bibitem{ras} Y. A. Bychkov and E. I. Rashba, J. Phys. C {\bf 17}, 6039
  (1984).
\bibitem{dress} G. Dresselhaus, Phys. Rev. {\bf 100}, 580 (1955).
\bibitem{burkov}A. A. Burkov, A. S. N\'{u}\~{n}ez, and A. H. McDonald, Phys. Rev. B {\bf 70}, 155308 (2004).
\bibitem{loss1}J. Schliemann and D. Loss, Phys. Rev. B {\bf 69}, 165315 (2004).
\bibitem{mish}E. G. Mishchenko, A. V. Shytov, and B. I. Halperin, Phys. Rev. Lett. {\bf 93}, 226602 (2004).
\bibitem{inoue}J. I. Inoue, G. E. W. Bauer, and L. W. Molenkamp, Phys. Rev. B {\bf 67}, 033104 (2003);
{\em ibid} {\bf 70}, 041303 (2004).
\bibitem{sheng1}D. N. Sheng, L. Sheng, Z. Y. Weng, and F. D. M. Haldane, cond-mat/0504218.
\bibitem{lei}S. Y. Liu and X. L. Lei, cond-mat/0411629.
\bibitem{sheng}L. Sheng, D. N. Sheng, and C. S. Ting, Phys. Rev. Lett. {\bf 94}, 016602 (2005).
\bibitem{nikolic}B. K. Nikoli\'{c}, L. Z\^{a}rbo, and
S. Souma, cond-mat/0408693.
\bibitem{pareek}T. P. Pareek, Phys. Rev. Lett. {\bf 92}, 76601 (2004).
\bibitem{murakami}S. Murakami, Phys. Rev. B {\bf 69}, 241202 (2004).
\bibitem{chen}W. Q. Chen, Z. Y. Weng, and D. N. Sheng, cond-mat/0502570.
\bibitem{loss}J. Schliemann and D. Loss, Phys. Rev. B {\bf 71}, 085308 (2005).
\bibitem{zhang}B. A. Bernevig and S. C. Zhang, cond-mat/0411457.
\bibitem{trebin}H.R. Trebin, U. R\"{o}ssler, and R. Ranvaud,
Phys. Rev. B {\bf 20}, 686 (1979).
\bibitem{strain}G. L. Bir and G. E. Pikus, {\em Symmetry and Strain-Induced Effects
in Semiconductors} (Wiley, New York, 1974).
\bibitem{Da} S. Datta, {\em Electronic Transport in Mesoscopic
    Systems} (Cambridge University Press, New York, 1995).
\bibitem{Bu} M. B{{\"u}}ttiker, Phys. Rev. Lett. {\bf 57}, 1761 (1986).
\bibitem{nitta}J. Nitta, T. Akazaki, H. Takayanagi, and T. Enoki, Phys. Rev. Lett. {\bf 78}, 1335 (1996).
\end{thebibliography}
\end{document}